# Electron transport and optical properties of shallow GaAs/InGaAs/GaAs quantum wells with a thin central AlAs barrier

V A Kulbachinskii[1,*], I S Vasil'evskii[1,3], R A Lunin[1], G Galistu[2], A de Visser[2], G B Galiev[3], S S Shirokov[3] and V G Mokerov[3]

[1] *Low Temperature Physics Department, Moscow State University, 119992, GSP-2, Moscow, Russia,*
[2] *Van der Waals-Zeeman Institute, University of Amsterdam, Valckenierstraat 65, 1018 XE Amsterdam, The Netherlands,*
[3] *Institute of UHF Semiconductor Electronics, 117105, Moscow, Russia*

**Abstract -** Shallow GaAs/InGaAs/GaAs quantum well structures with and without a three monolayer thick AlAs central barrier have been investigated for different well widths and Si doping levels. The transport parameters are determined by resistivity measurements in the temperature range 4-300 K and magnetotransport in magnetic fields up to 12 T. The (subband) carrier concentrations and mobilities are extracted from the Hall data and Shubnikov-de Haas oscillations. We find that the transport parameters are strongly affected by the insertion of the AlAs central barrier. Photoluminescence spectra, measured at 77 K, show an increase of the transition energies upon insertion of the barrier. The transport and optical data are analyzed with help of self-consistent calculations of the subband structure and envelope wave functions. Insertion of the AlAs central barrier changes the spatial distribution of the electron wave functions and leads to the formation of hybrid states, i.e. states which extend over the InGaAs and the delta-doped layer quantum wells.

*Corresponding author:
V.A. Kulbachinskii, Low Temperature Physics Department, Physics Faculty, Moscow State University, 119992, GSP-2, Moscow, Russia. (e-mail: kulb@mig.phys.msu.ru).

## 1. Introduction

In the past fifteen years research into quantum well (QW) structures with thin barriers has attracted considerable interest. Optical studies on e.g. GaAs QWs with thin AlAs or $Al_{1-x}Ga_xAs$ barrier layers incorporated in the well region demonstrated that the energy spectrum of the two-dimensional electrons could be tuned by changing either the barrier thickness or its height [1,2]. Such a tuning might be utilized for instance in infrared photodetectors or lasers [3].

For practical use of quantum well structures high electron mobilities are desirable, and therefore it is of much interest to suppress electron-phonon scattering, which is dominant in modulation doped quantum well structures at temperatures above 100 K. This might be accomplished by inserting a thin barrier which acts as a phonon wall. For instance in transport experiments reported in Ref.4 an increase in the electron mobility was observed when three AlAs barriers were inserted into a GaAs/AlAs multiple QW [4]. The reduction in scattering rate was attributed to the confinement of optical phonons [4], but in a theoretical paper [5] another explanation of the effect was suggested, namely a modulation of electron states. In several theoretical papers [6-9] it has been calculated that the introduction of thin AlAs barriers in rectangular QWs leads to suppression of intersubband scattering by optical phonons, which in turn enhances the electron mobility. Other theoretical work has argued against an observable enhancement of the mobility [10,11]. Clearly, consensus is lacking.

Surprisingly, until today no systematic transport studies have been undertaken for the case of a simple structure with a single barrier incorporated in the QW. The majority of the experimental work is devoted to the investigation of optical properties and subband formation in complex structures, e.g. heavily doped pseudomorphic high electron mobility transistors [12]. In this work we focus on electron transport properties of shallow InGaAs QW structures with a thin AlAs barrier incorporated in the center of the QW. We investigate how the transport parameters depend on the doping level and QW width.

## 2. Samples

Pseudomorphic $In_{0.12}Ga_{0.88}As$ quantum wells with and without AlAs barrier were grown by molecular-beam epitaxy on semi-insulating (001) GaAs substrates. The structures are schematically shown in Fig.1. The QW samples consist of the following layers: a GaAs buffer layer 0.6 μm thick, a Si δ-doping layer, a GaAs spacer layer 8.5 nm thick, the $In_{0.12}Ga_{0.88}As$ quantum well with well widths $L_{QW}$ of 8 or 12 nm, a GaAs spacer layer 8.5 nm thick, an upper Si δ-doping layer, and an i-GaAs layer 75 nm thick. The latter was grown in order to eliminate surface potential effects. The structures were covered with a cap layer of Si-doped GaAs 10 nm thick. The substrate temperature was 510 °C for the pseudomorphic QW and 590 °C for the other layers. Samples were prepared with δ-doping layers with Si concentrations of $3.2 \times 10^{12}$ cm$^{-2}$ (heavily doped, samples #1 and #2) and $\sim 1 \times 10^{12}$ cm$^{-2}$ (moderately doped, samples #3 - #6). Samples without (#1, #3, #5) and with barrier (#2, #4, #6) were prepared. The barrier consists of three monolayers of AlAs grown in the center of the QW. The growth was interrupted for 30 s before and after depositing the QW and barrier layers. Sample pairs (#1, #2), (#3, #4) and (#5, #6) were prepared within the same growth cycle and differ by the barrier layer only. The structural and electro-physical characterization of the samples has been reported in Ref. [13]. In order to carry out transport measurements all samples were prepared in Hall bar geometry by conventional lithography and wet etching. In order to attach current and voltage leads, AuGe/Ni/Au ohmic contact pads were made on the samples.

## 3. Transport properties

The temperature dependence of the sheet resistance measured for $T = 4.2$-300 K is shown in Fig. 2 for all samples. For the heavily doped samples #1 and #2 the resistance attains lower values and has a weaker temperature variation than for the moderately doped samples (#3-#6). The single QW samples #3 and #5 show metallic behavior: i.e. the resistance decreases approximately linearly with decreasing temperature down to $\sim 70$ K, below which the resistance increases weakly. The temperature and magnetic field variation of the resistance below 70 K has been studied in detail and can be attributed to weak localization effects [14].

The insertion of the barrier has a pronounced effect on the sheet resistance, notably in the moderately doped samples, although the barrier is quite thin. In samples #4 and #6 the value of the resistance at $T= 4.2$ K increases by a factor 3 and 7 compared to samples #3 and #5, respectively. The large difference in resistance due to insertion of the barrier decreases when the temperature increases. The resistance values of the single QW sample #5 are smaller than those of sample #3, although the well width is smaller ($L_{QW}$ = 8 nm compared to 12 nm). This is due to the slightly larger carrier concentration in sample #5 ($\sim 5$%) as was determined by the low-temperature Hall data (see below).

Electron Hall densities $n_H$ and Hall mobilities $\mu_H$ were determined at temperatures of 4.2, 77 and 300 K for all samples. An overview of the results is presented in Table 1. For the heavily doped samples #1 and #2 the Hall density amounts to 2.6-2.7x10$^{12}$ cm$^{-2}$ and is roughly temperature independent (to within $\sim 10$%). Also the mobility is quite low, which indicates that ionized impurity scattering is dominant. For the moderately doped single QWs (samples #3 and #5) the temperature variation of $n_H$ and $\mu_H$ is consistent with the metallic behavior observed in the resistance. The overall increase of the mobility with decreasing temperature is attributed to the reduction in phonon scattering rate. However, in the samples with barrier, #4 and #6, the Hall mobility on the whole decreases with decreasing temperature. Interestingly, at low temperatures (4.2 K and 77 K) the insertion of the barrier leads to a strong reduction of mobility by a factor 3-5, although the Hall density is roughly constant or even shows an increases (< 20%).

**4. Shubnikov-de Haas oscillations and quantum Hall effect**

The longitudinal $R_{xx}$ and Hall resistance $R_{xy}$ was measured for all samples in magnetic fields $B$ up to 12 T in the temperature range 0.25-4.2K. Typical results obtained at $T = 0.25$ K are shown in Fig.3 for samples #1, #2, #5 and #6. The overall behavior (non-oscillatory component) of the high-field magnetoresistance for the heavily and moderately doped samples is quite different: while for samples #1 and #2 the magnetoresistance has positive quadratic field dependence, for samples #3-#6 only a (initially sharp) negative magnetoresistance is observed, which is indicative of weak localization in low-density two-dimensional semiconductor structures. Superposed on the monotonous component, the longitudinal resistance shows pronounced Shubnikov-de Haas (SdH) oscillations. In Fig.4 we present the Fast Fourier transforms of the $R_{xx}(1/B)$ dependencies of the SdH signals, where we have scaled the frequency axis to yield the two-dimensional electron densities. For samples #1-#5 one main frequency peak is found, which indicates the presence of at least one occupied high-mobility subband. For sample #2 a shoulder is visible in the Fourier transform, which indicates the occupation of a second subband. For sample #6 no clear frequency can be detected in the FFT, which is due to the low mobility (see Table 1) and the long oscillation period which extends into the quantum Hall regime. The resulting values for the SdH density, $n_{SdH}$, are collected in Table 2, together with the quantum mobilities, determined from the envelope of the SdH oscillations [15]. In the heavily doped samples $n_{SdH}$ decreases from $1.35 \times 10^{12}$ cm$^{-2}$ to $\sim 0.6 \times 10^{12}$ cm$^{-2}$ when the barrier is inserted. The SdH densities are much lower than the Hall density, indicating that several subbands with different electron mobilities are populated. For the moderately doped samples the SdH and Hall densities are all of the same order (the Hall densities being 10-20% higher), which indicates that transport is dominated by a high-mobility subband. Note that the barrier insertion weakly decreases the carrier concentration in the moderately doped samples.

The transverse resistance $R_{xy}$ for samples #3-#6 shows the quantum Hall effect (QHE). The QHE is most pronounced in the samples without barrier, because of the higher mobility. For samples #3 and #5 clear integer plateaus are observed at $T=0.25$ K for non-spin split Landau levels with filling factors $\nu = 4$ and $\nu = 2$ (see Fig.3b). At the integer filling factors $R_{xx} = 0$, which demonstrates the absence of parallel conduction. In samples #1 and #2 parallel conduction due to the population of several subbands hampers the observation of the QHE.

## 5. Photoluminescence

The photoluminescence (PL) spectra of all samples have been measures at $T$= 77 K. The results are reported in Fig. 5. All PL spectra of the $In_{0.12}Ga_{0.88}As$ quantum wells exhibit a pronounced maximum in the energy range 1.35-1.47 eV, which is somewhat below the transition in bulk GaAs at 1.508 eV. For the single QWs (#1, #3 and #5) the peaks are relatively broad and the PL intensity rise differs for the different sample, indicating the presence of several transition energies. For samples with barrier (#2, #4 and #6), the PL peaks are less broad, which indicates that the electron levels are more closely spaced. A most important observation is that incorporating the barrier leads to a significant upward shift of the spectra of the order of 0.06 eV, without a substantial decrease of PL intensity. We also note that for the single QW samples #1 and #3, which have the same well width $L_{QW}$= 12 nm but different doping levels, the transition energies differ slightly (by 0.02 eV). However, upon insertion of the barrier (#2 and #4) this energy difference disappears.

## 6. Subband structure and wave functions

The conduction band profile and the subband structure were calculated for all the structures by solving the Schrödinger and Poisson equations self consistently (see e.g. Ref.16). In order to achieve adequate modeling of the δ-doped layer we used a finite distribution width of 5 nm, which is the characteristic width of the Si δ-layer in GaAs at the applied growth temperature [17]. The conduction band profiles for the heavily doped samples #1 and #2 and the moderately doped samples #3 and #4 are qualitatively different as shown in Fig.6, where we have taken the Fermi level as zero energy reference. In the heavily doped samples the δ-doped layers form additional quantum wells almost symmetrically bordering the $In_{0.12}Ga_{0.88}As$ QW. In sample #1 the envelope wave function $\Psi_0$ of the ground state (energy $E_0$) is predominantly situated in the $In_{0.12}Ga_{0.88}As$ QW, but partially penetrates in the V-shaped δ-layer QWs. The wave functions $\Psi_1$ and $\Psi_2$ are mainly confined in the δ-layer QWs and result in two splitted subbands labeled $E_1$ and $E_2$. The calculated values of the subband electron concentration $n_i$ (where $i$ is the subband index) are reported in Table 2. Insertion of the AlAs barrier (sample #2) results in a tunnel splitting of the central QW state into the wave functions $\Psi_2$ and $\Psi_3$ and an upward shift of the subband energies, now labeled $E_2$, $E_3$. Hence the central QW states are no longer the ground state of the whole system. The wave functions $\Psi_2$ and $\Psi_3$ are strongly reconstructed compared to $\Psi_0$ in sample #1, and rather form a hybrid state in the quantum well and δ-doped regions. The wave functions in the δ-doped wells (now $\Psi_0$ and $\Psi_1$) are less affected by the insertion of the barrier. The electron density increases in the region of the δ-doped wells and decreases in the QW region upon insertion of the barrier (see Table 2). This is due to the relatively shallow central QW. Also band bending considerably affects the band structure. The formation of a hybrid state due to heavy doping has also been reported for InGaAs quantum wells [18].

In the moderately doped samples #3 and #5 the V-shaped δ-layer QWs are significantly weaker and band bending remains relatively small. Moreover, the conduction band profile is asymmetric: only one non-QW state $\Psi_1$ with energy $E_1$ forms below $E_F$ in the δ-layer. This state is located in the lower δ-layer QW (i.e. at ~ 115 nm below the surface). The electron concentration $n_1$ in this subband is however negligible at low temperatures. The energy difference between the ground state energy level $E_0$ and the level $E_1$ is high. Hence, the electrons in the ground state subband of the single QW are largely confined in the QW ($\Psi_0$ for the sample #3, see Fig. 6). Just as in the case for the heavily doped samples, insertion of the barrier into the QW leads to a significant redistribution of the wave function towards the δ-doped region ($\Psi_0$ for sample #4). The energy level $E_0$ (and also $E_F$) shift up with respect to the QW bottom. $E_1$ increases due to Fermi level increase, and an associated electron concentration $n_1$ results from the calculations (see Table 2).

**7. Discussion**

In the simple case of a single quantum well one expects that insertion of a central barrier leads to an increase of the energy levels and a decrease of the electron densities in the occupied subbands. The photoluminescence data are consistent with this idea and reveal an overall energy shift of ~ 0.06 eV. However, the transport measurements show that the situation is more complicated. In the moderately doped samples, e.g. sample #3, $n_{SdH}= 0.45 \times 10^{-12}$ cm$^{-2}$, which decreases only slightly upon insertion of the barrier (sample #4), i.e. less than ~ 10%. This is explained by the Fermi level shifting up with respect to the QW bottom, when the barrier is inserted. This causes the energy separation between the hole and electron bands to increase considerably, while the difference $E_F–E_i$ remains small (~ 2 meV). It is this latter energy difference which determines the carrier concentration. For the heavily doped samples (#1 and #2) the transport measurements show that the difference $E_F–E_i$ indeed changes: for sample #1 $n_{SdH}$ of the high-mobility subband associated with the central QW is $1.35 \times 10^{-12}$ cm$^{-2}$, which decreases to ~$0.6 \times 10^{-12}$ cm$^{-2}$ in sample #2.

The energy band structure calculations are most useful for clarification of the transport results as they reveal the strong influence of the V-shaped δ-layer QWs. The key feature is the delocalization of the central wave function into the δ-layer region upon insertion of the barrier (hybrid state), which has its origin in the shallowness of the central QW and a relatively strong δ-doping level. To account for the unusual transport behavior, the quantum $\mu_q$ and transport $\mu_t$ mobilities due to ionized impurities scattering in the various subbands have been calculated for the various subbands [19]. The results are collected in Table 2.

Let us first consider the moderately doped samples, e.g. #3 and #4. The calculations show that while the barrier is added at the QW center, the ground state energy level $E_0$ shifts up with respect to the QW bottom (+22 meV), so its wave function $\Psi_0$ has now noticeable amplitude in the lower δ-layer (i.e. at 115 nm) QW. The decrease of the Hall mobility at all temperatures is naturally explained by the additional scattering contribution due to ionized impurities in the δ-layer regions when the hybrid state is formed. As follows from the data in Table 2, the calculated transport mobility $\mu_t$ is high in the single QW samples #3 and #5, and much smaller in the QW samples #4 and #6 with barrier. This confirms that in the latter samples ionized impurity scattering dominates. The decrease of the mobility is most pronounced in sample #6, i.e the sample with small $L_{QW}$= 8 nm, because the wave function $\Psi_0$ has the strongest amplitude in the δ-layer region.

In the heavily doped samples comparison of the SdH and Hall concentrations at $T$= 4.2 K reveals that the QW subband is occupied by slightly less electrons ($n_{SdH}$= 1.35x10$^{12}$ cm$^{-2}$, calculated value $n_i$=1.3 x10$^{12}$) than the subbands in the δ-layer potential wells (with a total electron density ~ 1.4-1.5x10$^{12}$ cm$^{-2}$). The insertion of the barrier (sample #2) effectively shifts the QW subband upwards (from $E_0$ to $E_2$ and $E_3$, i.e. an energy shift $\Delta E$ ~ 20 meV), while $E_F$ is "stabilized" by the high electron concentration in the δ-layer QWs. Thus the hybrid wave functions $\Psi_2$ and $\Psi_3$ are now the central QW states. The mobility calculations show that the highest value (see underlined values in Table 2) is obtained for the third subband with wave function $\Psi_2$. The observed SdH oscillation in sample #2 is attributed to the lowest hybrid QW state $\Psi_2$ and is consistent with the calculated value $n_2$. The mobility in this subband is still high, because i) $|\Psi_2|^2$ is small in the δ-layer area and ii) the electrons in the δ-layer QWs effectively screen the ionized impurity potential.

## 8. Summary

The transport and optical properties of shallow GaAs/In$_{0.12}$Ga$_{0.88}$As/GaAs quantum wells with and without a three monolayer thick central AlAs barrier have been investigated. Magnetotransport and photoluminescence measurements were performed on samples prepared with two different quantum well widths and different Si doping levels. The PL data show an overall shift of the spectra to higher transition energies ($\Delta E$~ 0.05 eV), while the electron concentration extracted from the Hall data decreases only slightly. The mobility decreases upon insertion of the central barrier. Self-consistent calculations of the subband structure and envelope wave functions reveal a strong influence of the δ-doping regions on the conduction band profile: additional V-shaped quantum wells are formed in the δ-doping regions. Consequently, the central QW wave function extends into the δ-doping regions and forms a "hybrid" wave function. The hybrid character becomes more pronounced when the central barrier is incorporated in the structures and accordingly the electron density displaces towards the δ-layers. This results in a change of the dominant scattering mechanism from phonon to ionized impurity scattering.

**Acknowledgements**

This work was supported by Russian Foundation for Basic Research, grant 05-02-17029-a, and FOM (Dutch Foundation for Fundamental Research of Matter).

**Tables and table captions**

**Table 1** Structural and transport parameters (at $T$ = 300 K, 77 K and 4.2 K) of the InGaAs QW samples.

| # | $L_{QW}$ nm | $N_d$ (Si) $10^{12}$ cm$^{-2}$ | $T$=300 K | | $T$=77 K | | $T$=4.2 K | |
|---|---|---|---|---|---|---|---|---|
| | | | $n_H$ $10^{12}$ cm$^{-2}$ | $\mu_H$, cm$^2$/Vs | $n_H$, $10^{12}$ cm$^{-2}$ | $\mu_H$ cm$^2$/Vs | $n_H$ $10^{12}$ cm$^{-2}$ | $\mu_H$ cm$^2$/Vs |
| 1 | 12 | 3.2 | 2.72 | 3830 | 3.0 | 4700 | 2.86 | 3 800 |
| 2 | 12+ b* | 3.2 | 2.6 | 3150 | 2.33 | 5420 | 2.61 | 3 300 |
| 3 | 12 | 1.04 | 0.54 | 5740 | 0.79 | 18500 | 0.52 | 10 000 |
| 4 | 12+ b* | 1.04 | 0.42 | 4810 | 0.78 | 5300 | 0.57 | 2 070 |
| 5 | 8 | 1.1 | 0.53 | 5910 | 0.76 | 18700 | 0.59 | 7 980 |
| 6 | 8 + b* | 1.1 | 0.50 | 4000 | 0.87 | 3570 | 0.47 | 1 520 |

\* +b indicates samples with inserted central AlAs barrier (1 nm)

**Table 2** Transport parameters of the InGaAs QW samples #1-#6. $n_H$ is the Hall and $n_{SdH}$ the Shubnikov-de Haas electron concentration measured at $T$ = 4.2 K, $\mu_{SdH}$ is the quantum mobility determined from the SdH effect, $n_i$ is the calculated subband concentration, and $\mu_{qi}$ and $\mu_{ti}$ the calculated quantum and transport mobility, respectively, for subband $i$, due to ionized impurity scattering.

| Sample # | $L_{QW}$ (nm) | barrier AlAs (3ML) | $n_{SdH}$ | $n_i$ (i) | $n_H$ | $\mu_{SdH}$ | $\mu_{qi}$ | $\mu_{ti}$ |
|---|---|---|---|---|---|---|---|---|
| | | | | $10^{12}$ cm$^{-2}$ | | | cm$^2$/Vs | |
| 1 | 12 | – | 1.35 | 1.30 (0) | 2.86 | 2 700 | 7 780 | 15 800 |
| | | | | 0.83 (1) | | | 640 | 2 870 |
| | | | | 0.61 (2) | | | 470 | 2 160 |
| 2 | 12 | + | – | 1.05 (0) | 2.61 | 1 660 | 880 | 3 150 |
| | | | – | 0.97 (1) | | | 890 | 2 330 |
| | | | 0.66 | 0.65 (2) | | | 2 090 | 4 600 |
| | | | 0.58 | 0.52 (3) | | | 940 | 2 500 |
| 3 | 8 | – | 0.49 | 0.51 (0) | 0.52 | 1 400 | 3 210 | 33 700 |
| 4 | 8 | + | 0.44 | 0.44 (0) | 0.57 | 920 | 2 770 | 6 470 |
| | | | | 0.1 (1) | | | | |
| 5 | 8 | – | 0.55 | 0.48 (0) | 0.59 | 1 430 | 2 850 | 28 800 |
| 6 | 8 | + | – | 0.43 (0) | 0.47 | – | 1 650 | 4 000 |
| | | | | 0.13 (1) | | | | |

**Figure captions**

Fig. 1.  Schematic sample structure for the $In_{0.12}Ga_{0.88}As$ quantum wells:
(a) without and (b) with an AlAs central barrier.

Fig. 2.  Temperature variation of the sheet resistance for the $In_{0.12}Ga_{0.88}As$ quantum wells with (#2,#4,#6) and without (#1,#3,#5) the AlAs central barrier.

Fig. 3.  Longitudinal $\rho_{xx}$ and transverse $\rho_{xy}$ resistivity measured at $T$=0.25 K for (a) samples #1 and #2 and (b) for samples #5 and #6.

Fig. 4.  Fast Fourier spectra of the Shubnikov-de Haas data observed for samples #1-#5 at T=0.25 K. Notice the horizontal axis yields the SdH electron density.

Fig. 5.  Photoluminescence spectra measured at 77 K for the $In_{0.12}Ga_{0.88}As$ quantum wells with (#2,#4,#6) and without (#1,#3,#5) the AlAs central barrier.

Fig. 6.  Calculated conduction band profiles, electron wave functions and subband energy levels for $In_{0.12}Ga_{0.88}As$ quantum wells with (#2,#4) and without (#1,#3) an AlAs central barrier.

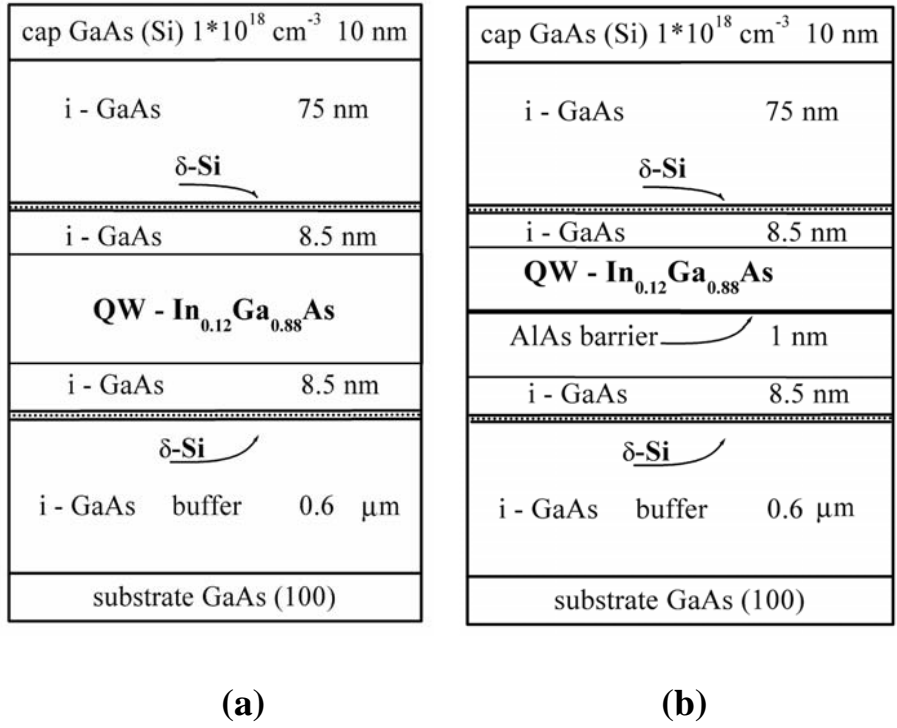

**(a)** **(b)**

**Fig. 1**

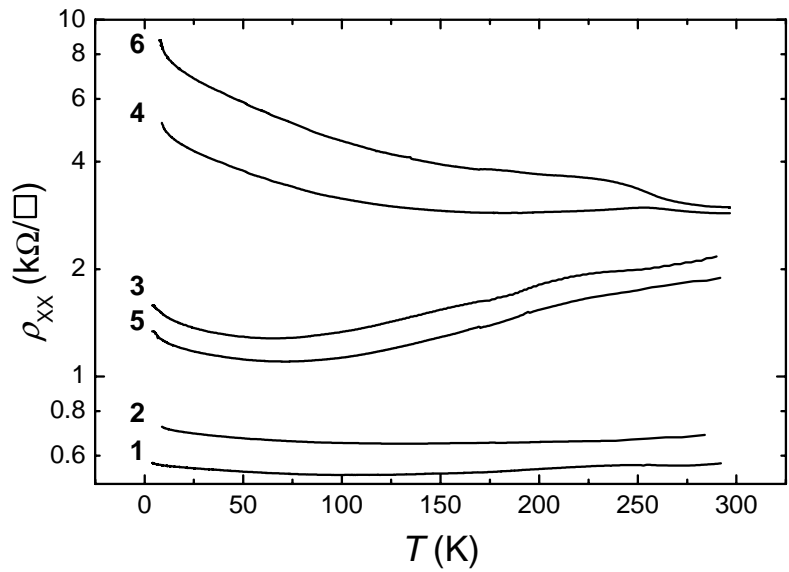

**Fig. 2**

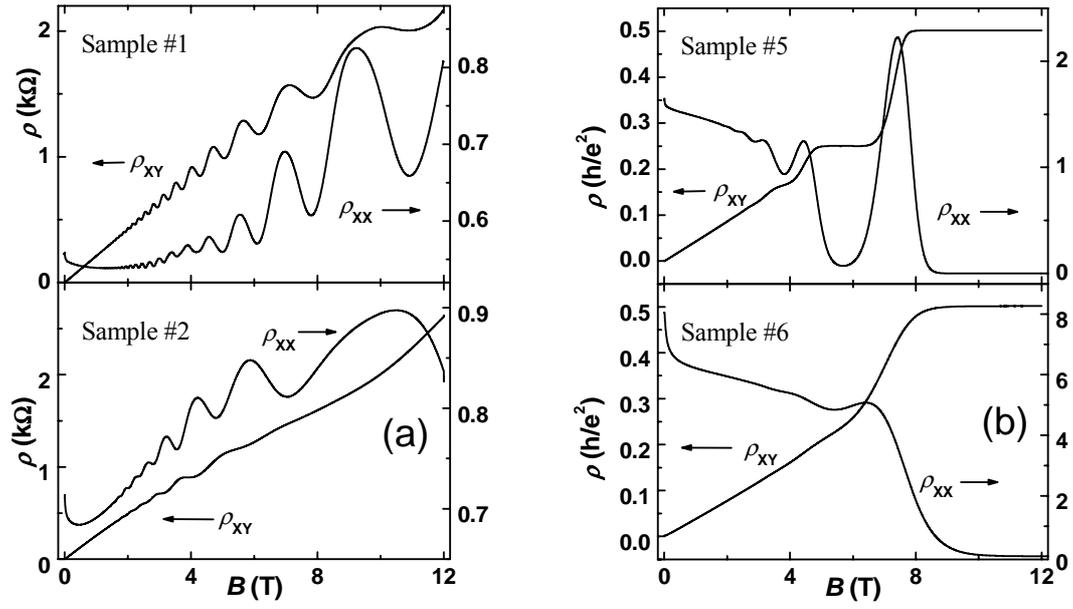

**Fig. 3**

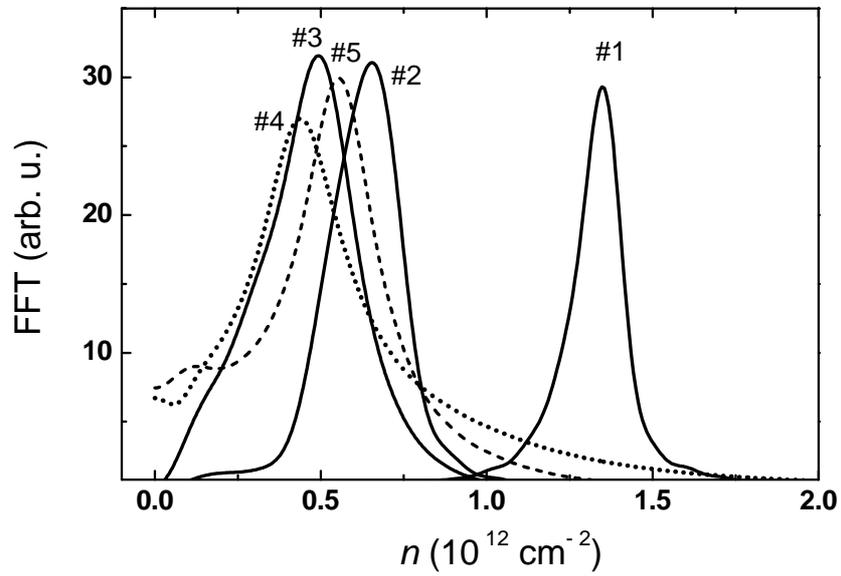

**Fig. 4**

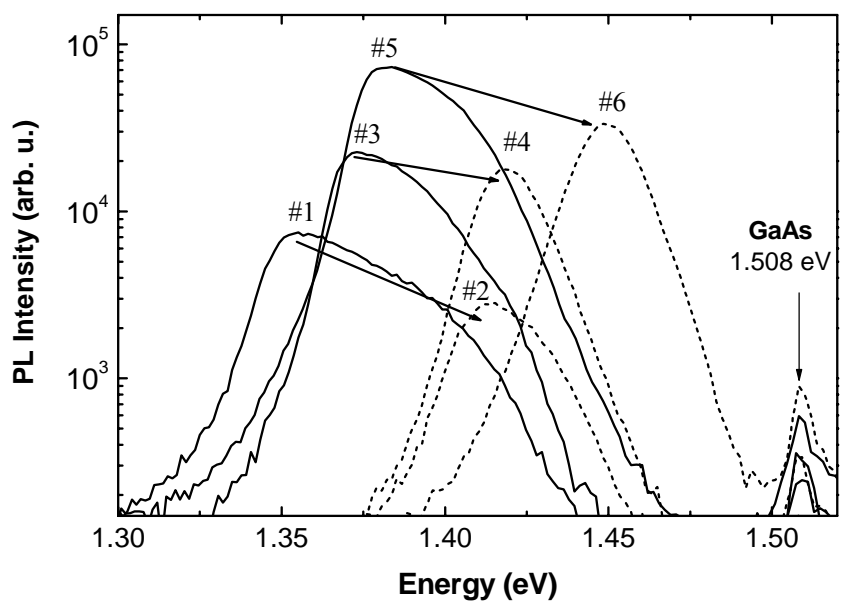

**Fig. 5**

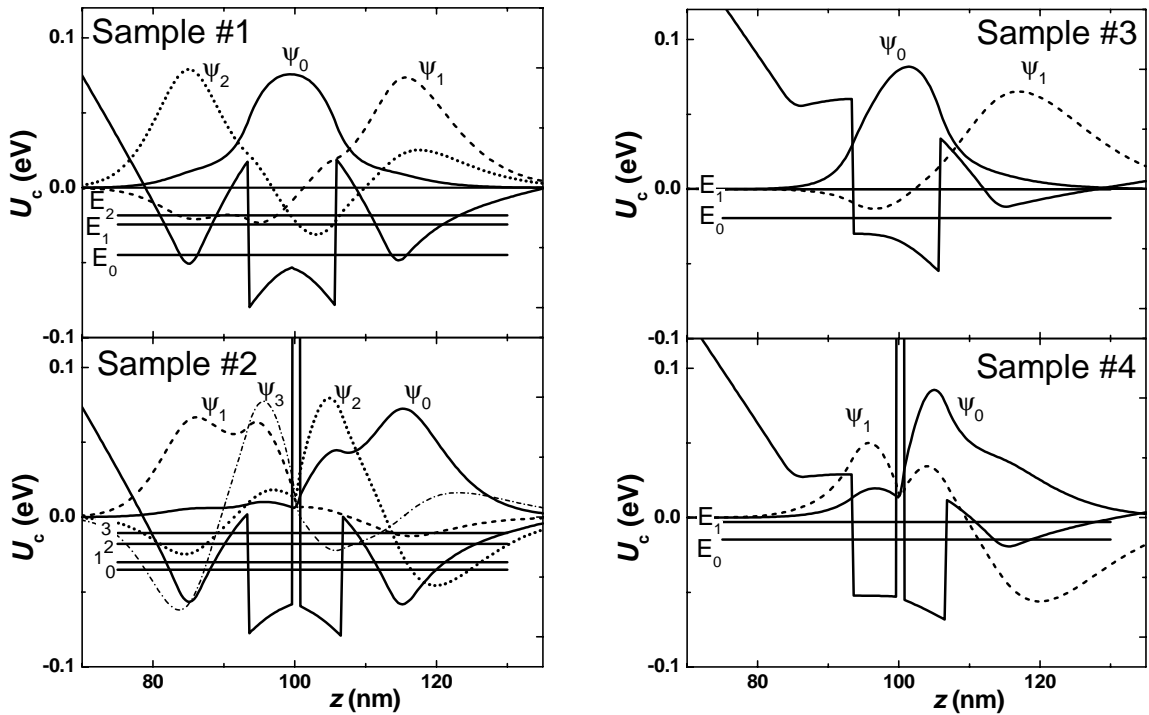

**Fig. 6**